\documentclass{osa-article}

\journal{osajournal}

\articletype{Research Article}

\begin{document}

\title{Motional $N$-phonon Bundle States of A Trapped Atom with Clock Transitions}

\author{Yuangang Deng,\authormark{1,*} Tao Shi,\authormark{2,3} Su Yi,\authormark{2,3,4,\dag}}

\address{\authormark{1}Guangdong Provincial Key Laboratory of Quantum Metrology and Sensing $\&$ School of Physics and Astronomy, Sun Yat-Sen University (Zhuhai Campus), Zhuhai 519082, China\\
\authormark{2}CAS Key Laboratory of Theoretical Physics, Institute of Theoretical Physics, Chinese Academy of Sciences, Beijing 100190, China\\
\authormark{3}CAS Center for Excellence in Topological Quantum Computation, University of Chinese Academy of Sciences, Beijing 100049, China\\
\authormark{4}School of Physical Sciences, University of Chinese Academy of Sciences, Beijing 100049, China}

\email{\authormark{*}dengyg3@mail.sysu.edu.cn}
\email{\authormark{*}syi@itp.ac.cn}


\begin{abstract}
Quantum manipulation of individual phonons could offer new resources for studying fundamental physics and creating an innovative platform in quantum information science. Here, we propose to generate quantum states of strongly correlated phonon bundles associated with the motion of a trapped atom. Our scheme operates in the atom-phonon resonance regime where the energy spectrum exhibits strong anharmonicity such that energy eigenstates with different phonon numbers can be well-resolved in the parameter space. Compared to earlier schemes operating in the far dispersive regime, the bundle states generated here contain a large steady-state phonon number. Therefore, the proposed system  can be used as a high quality multiphonon source. Our results open up the possibility of using long-lived motional phonons as quantum resources, which could provide a broad physics community for applications in quantum metrology.
\end{abstract}

\section{Introduction}
Engineering special nonclassical quantum states are of paramount importance in quantum information science, metrology, and exploring fundamental physics~\cite{duan2001long,PhysRevLett.96.010401,kok2007linear,pezze2018quantum,braun2018quantum}. Especially, $n$-photon states play an essential role in a wide range of quantum technologies, including high-NOON states~\cite{afek2010high}, quantum communication~\cite{kimble2008quantum}, lithography~\cite{PhysRevLett.87.013602}, spectroscopy~\cite{PhysRevLett.115.196402,RevModPhys.88.045008}, and biological sensing~\cite{denk1990two,horton2013vivo}. Methods for generating $n$-photon states were proposed theoretically in cavity quantum electrodynamics (QED)~\cite{munoz2014emitters,munoz2018filtering,PhysRevLett.117.203602}, Rydberg atomic ensembles~\cite{bienias2014scattering,maghrebi2015coulomb}, and atom-coupled photonic waveguides~\cite{PhysRevLett.115.163603,PhysRevX.6.031017,PhysRevLett.118.213601}. Among them, $n$-photon bundle states generated through either Mollow physics~\cite{munoz2014emitters,munoz2018filtering} or deterministic parametric down-conversion~\cite{PhysRevLett.117.203602} are of particular interest, since they possess special statistic properties. However, due to the intrinsic weak scattering interactions between photons, the experimental realization of $n$-photon states still remains a challenge.

On the other hand, the ability to manipulate individual phonon allows the experimental creation of $n$-phonon states in both circuit quantum acoustodynamics~\cite{moores2018cavity,PhysRevX.9.021056} and macroscopic mechanical resonators~\cite{Arrangoiz-Arriola,manenti2017circuit,chu2018creation}. These low-energy and long-lived novel phonon states could facilitate the study of the decoherence mechanisms~\cite{Markus14} and the building of the quantum memories and transducers~\cite{PhysRevX.5.031031,PhysRevLett.119.180505,PhysRevX.8.031007}. More interestingly, there were was also proposed that $n$-phonon bundle states can be generated in acoustic cavity QED by employing the Stoke processes~\cite{PhysRevLett.124.053601} and hybrid system of nitrogen-vacancy centers and nanomechanical resonators via sideband engineering~\cite{PhysRevA.100.043825}.

We note that existing schemes~\cite{munoz2014emitters,munoz2018filtering,PhysRevLett.117.203602,PhysRevLett.124.053601} for creating $n$-quanta bundle states operate in the far dispersive regime where the frequency of the photonic/phononic mode is far detuned from the transition frequency of the two-level system. Hence the resulting steady-state photon/phonon numbers are typically very small. Moreover, in order to resolve states with distinct phonon numbers, many schemes~\cite{moores2018cavity,PhysRevX.9.021056,manenti2017circuit,chu2018creation,Arrangoiz-Arriola,PhysRevLett.124.053601} also require that the two-level system is strongly coupled to the phononic modes, which demands high-finesse acoustic cavities or mechanical resonators with long coherence time and poses a challenge to the current experiments.

In this work, we propose to generate $n$-phonon bundle states by utilizing the motional degrees of freedom of a trapped alkaline-earth atom. A position-dependent clock laser is introduced to couple atom's center-of-mass motion to its electronic ground and the long-lived excited states, which leads to a generalized quantum Rabi model (QRM) with unprecedented tunability. We then investigate the $n$-phonon bundle states prepared in the resonant regime where the frequency of the motional mode is in resonance with two-level atom. We show that due to the strong anharmonicity of the energy spectrum, distinct motional $n$-phonon bundle states can be well resolved in the atom-phonon resonance regime. Compared to the existing schemes for generating $n$-quanta bundle states, the system proposed here has following advantages: i) Since our scheme operates in the resonant regime, the typical average steady-state phonon number is much larger than that of the schemes operating in the far dispersive regime. ii) Strong coupling regime can be readily achieved here as both the motional state and the atomic internal states possess long lifetimes. iii) In our configuration, the effective pump field is provided by the high tunable clock detuning, our system also facilitates the study of Mollow physics without suffering heating and decoherence. Therefore, the proposed system can be used as a high-quality source for multiphonon states.

\section{Model and Hamiltonian} 
\begin{figure}[tbp]
	\includegraphics[width=0.98\columnwidth]{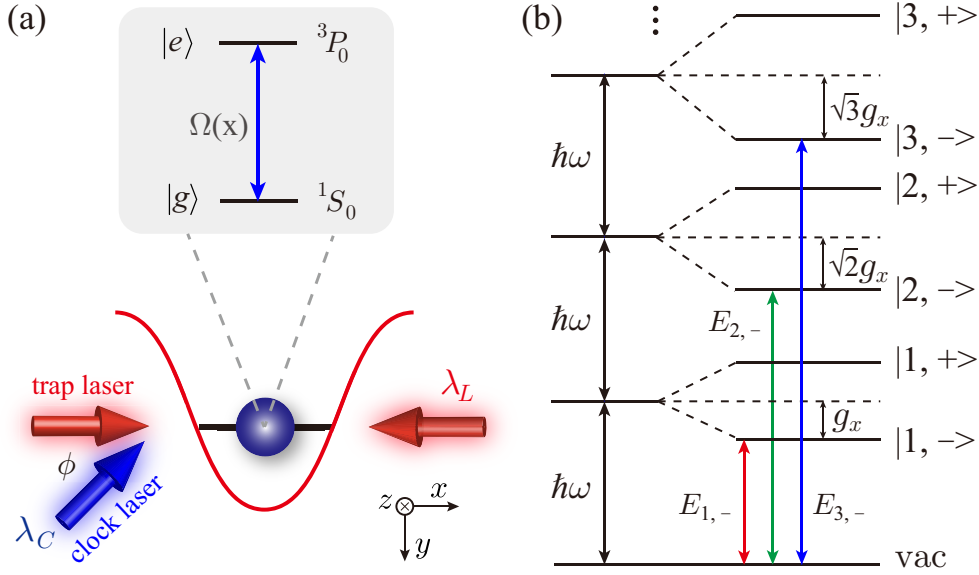}
	\caption{(a) Schematic of the system. A $^{87}$Sr atom is trapped in an one-dimensional potential formed by two counter propagating lasers of the `magic' wavelength. The $^1S_0-^3$$P_0$ transition of the atom is coupled by the clock laser at wavelength $\lambda_C$. The clock laser lies on the $xy$ plane making an angle $\phi$ to the trap lasers. (b) Anharmonic energy spectrum of the Hamiltonian $\mathcal{H}'$ with $g_k=g_x$ and $\Omega=\omega$. Here $|n,\pm\rangle$ denote the $n$th pair of dressed states with eigenenergies $E_{n,\pm} = \hbar (n\omega \pm g_x\sqrt{n})$.}
	\label{scheme}
\end{figure}

Without loss of generality, we consider a single ${}^{87}$Sr atom trapped in an one-dimensional (1D) harmonic potential along the $x$ direction. Figure~\ref{scheme}(a) illustrates the level structure and laser configuration of the system. An ultranarrow clock laser with wavelength $\lambda_C=698\,{\rm nm}$ drives the single-photon transition between the ground state ${}^1S_0$ ($|g\rangle$) and the long-lived excited state ${}^3P_0$ ($|e\rangle$) with the atomic clock transition frequency $\omega_a$. Since the lifetime of the excited state is roughly 160 seconds~\cite{kolkowitz2017spin}, the spontaneous emission and decoherence of this state can be safely ignored. We assume that the Rabi frequency of the clock laser takes a position-dependent form, $\Omega_l(x) = (\Omega_0x-i{\Omega}/{2}) e^{-i\kappa x}$, where $\Omega_0$ and $\Omega$ are coupling strengths and $\kappa=k_{C}\cos\phi$ is the effective laser wave vector that is tunable by varying the tilting angle $\phi$ of the clock laser. As shall be shown, the position-dependent Rabi coupling of $\Omega_l(x)$ is crucial to the success of our scheme and it can be experimentally generated by tailoring the clock laser by using a spatial light modulator~\cite{Beeler,Palima:07,Pasienski:08,PMID:23301155,Gaunt12}. We further assume that the 1D harmonic potential is state independent which can be generated by a trap laser at the `magic' wavelength $\lambda_L=813\,{\rm nm}$. After performing a gauge transformation $\left|g\right\rangle \rightarrow e^{i\kappa x/2}\left|g\right\rangle$ and $\left|e\right\rangle \rightarrow e^{-i\kappa x/2}\left|e\right\rangle$~\cite{PhysRevLett.108.125301}, the resulting Hamiltonian of the system is
\begin{align}
	{\cal {H}}/\hbar \!=\! \omega\hat{a}^\dag\hat{a} \!+\! \frac{\Omega}{2}\sigma_y \!+\! {\delta}\sigma_z\!+\! \frac{g_x}{2} (\hat{a}^\dag \!+\! \hat{a})\sigma_x \!+\!\! \frac{ig_{k}}{2}(\hat{a}^\dag \!\!-\!\hat{a})\sigma_z,
	\label{Rabi}
\end{align}
where $\omega$ is the frequency of the harmonic trap which, unlike the detuning of an optical cavity mode, is always positive, $\hat{a}$ is the annihilation operator of bosonic phonon for the external motional mode, $\sigma_{x,y,z}$ are the Pauli matrices, $\delta$ is the single-photon detuning from the bare clock transition, and $g_x=2\Omega_0 x_0$ and $g_k=\kappa\omega x_0$ with $x_0$ being the zero-point fluctuation amplitude of the harmonic oscillator. Here, $g_x$ and $g_k$ can be understood as the strengths of the spin-orbit coupling (SOC) in the real ($\sim x\sigma_x$) and momentum ($\sim p_x\sigma_z$) spaces~\cite{NATYJ2011SOCAT}, respectively.

To transfer Eq.~\eqref{Rabi} into a more familiar form, we introduce a spin rotation, $e^{{i\pi}\sigma_x/4}$, which rotates spin operators according to $\sigma_x\rightarrow \sigma_x$, $\sigma_y\rightarrow\sigma_z$, and $\sigma_z\rightarrow-\sigma_y$. As a result, Hamiltonian~\eqref{Rabi} becomes a generalized QRM,
\begin{align}
	{\cal {H}}'/\hbar \!=\! \omega\hat{a}^\dag\hat{a} \!+\! \frac{\Omega}{2}\sigma_z \!\!-\! {\delta}\sigma_y\!+\! \frac{g_{x}}{2} (\hat{a}^\dag \!\!+\! \hat{a})\sigma_x \!-\!\!\frac{ig_{k}}{2}(\hat{a}^\dag \!\!-\!\hat{a})\sigma_y,
	\label{Rabi1}
\end{align}
which is an essential building block in quantum information and quantum optics~\cite{rabi1936process,PhysRevLett.115.180404,liu2017universal,lv2018quantum}. Apparently, $\mathcal{H}'$ reduces to the standard QRM by setting $\delta=0$ and letting either $g_x$ or $g_k$ be zero. Furthermore, Hamiltonian~\eqref{Rabi1} has an extremely high tunability. For instance, it turns into the Jaynes-Cummings model (JCM) when $g_k=g_x$ and the anti-JCM when $g_k=-g_x$. Comparing with the experimental using the standard sideband transitions~\cite{lv2018quantum}, the emerged Hamiltonian of JCM or anti-JCM is exact without rotating-wave approximation, corresponding the high-frequency terms is completely eliminated in  Eq.~\eqref{Rabi1}. In particular, the ${\delta}\sigma_y$ term in Eq.~\eqref{Rabi1} which represents an effective classical pumping field of the atom is physically realized through the clock shift $\delta$. Consequently, the strong pumping regime ($\delta\gg \omega$) can be readily achieved without suffering severe laser induced heating, in contrast to the Raman induced SOC in atomic gases~\cite{Galitski2013}.

\section{Generalized quantum statistics}
For a complete description of the system, we should also take into account the dissipation of the phonons. As a result, the dynamics of the atom-phonon system is now described by the master equation
\begin{equation}
	\frac{d{\rho}}{dt} = -i [{\cal H}', {\rho}] + \frac{\kappa_e}{2} \mathcal
	{\cal{D}}[\sigma_-]\rho + \frac{\kappa_d}{2} \mathcal
	{\cal{D}}[\hat{a}]\rho + \gamma_d \mathcal{\cal{D}}[\hat{a}^\dag\hat{a}]\rho,
	\label{master}
\end{equation}
where $\rho$ is the density matrix of the atom-phonon system, $\gamma_e$ is atomic spontaneous emission rate of the clock state with $\sigma_-=(\sigma_x- i\sigma_y)/2$, $\kappa_d$ and $\gamma_d$ are, respectively, the decay rate and the dephasing factor of the phonons, and $\mathcal {D}[\hat{o}]\rho=2\hat{o} {\rho} \hat{o}^\dag - \hat{o}^\dag \hat{o}{\rho} - {\rho} \hat{o}^\dag \hat{o}$ denotes the Lindblad type of dissipation. For a given set of parameters, Eq.~\eqref{master} can be numerically evolved in the basis $\{|n,\sigma\rangle\}$ until a steady-state density matrix is obtained, where $|n\rangle$ is the Fock state of phonon and $|\sigma\rangle=|e\rangle$ or $|g\rangle$ represents the atomic state. We should emphasize that the realized effective Hamiltonian \eqref{Rabi1} of generalized QRM is originating from the rotating-wave approximation since $|g_{x,k}/\omega_a|< 10^{-10}$. Thus the standard Lindblad Eq.~(\ref{master}) is valid even the coupling strength $g_{x,k}$ comparable to the effective atom (phonon) detuning $\Omega$ ($\omega$)~\cite{walls2007quantum}. Therefore the dissipative terms of two-level atom and phonon emerged by external environment in our mode can be treated as independent with safely neglecting atom-phonon coupling~\cite{PhysRevA.84.043832,PhysRevLett.109.193602}.

To characterize the statistic properties of the phonons, we introduce the generalized $k$th-order correlation function,
\begin{align}
	g_n^{(k)}(\tau_1,\ldots,\tau_n)=\frac{\left\langle \prod_{i=1}^k\left[\hat{a}^{\dagger }(\tau_i)\right]^n \prod_{i=1}^k\left[\hat{a}(\tau_i)\right]^n\right\rangle}{\prod_{i=1}^k\left\langle \left[\hat{a}^{\dagger }(\tau_i)\right]^n\left[\hat{a}(\tau_i)\right]^n \right\rangle},
	\label{inequation}%
\end{align}
first introduced in Ref.~\cite{munoz2014emitters}, where $\tau_1\leq...\leq\tau_n$. As can be seen, this definition generalizes the standard $k$th-order correlation function~\cite{PhysRev.130.2529} for isolated phonons to bundles of $n$ phonons. We note that the equal time correlation function $g_n^{(k)}(0)$ for $\tau_1=\ldots=\tau_k$ is straightforwardly calculated using the steady-state density matrix solution of the Eq.~(\ref{master}); while the multitime correlation function $g_n^{(k)}(\tau_1,\ldots,\tau_k)$ with $\tau_1<\ldots<\tau_k$ can be obtained by utilizing the quantum regression theorem~\cite{carmichael2013statistical}. Now, an $n$-phonon bundle states ($n>1$) should satisfy two conditions: $g_1^{(2)}(0)>g_1^{(2)}(\tau)$ to ensure bunching with respect to single phonon and $g_n^{(2)}(0)<g_n^{(2)}(\tau)$ to secure antibunching between separated bundles of phonons~\cite{munoz2014emitters,PhysRevLett.117.203602}. We point out that the correlation functions of $g_1^{(2)}$ and $g_n^{(2)}$ can be directly extracted from the steady-state phonon-number distribution $p(q)=\mathrm{tr}(|q\rangle\langle q|\rho)$ which, for the motional phonons, can be measured via spin state-resolved projective measurement~\cite{lv2018quantum,wolf2019motional}. While for acoustic phonons, although $g_1^{(2)}$ which can be measured via Hanbury Brown and Twiss interferometer~\cite{birnbaum2005photon}, experimental measuring of the multiphonon $g_n^{(2)}$ is still a challenge task.

Finally, we specify the parameters used in numerical simulations. For ${}^{87}$Sr atom, the single-photon recoil energy $E_C/h= 4.68\,{\rm kHz}$ sets up the typical energy scale for the system, where $h$ is the Planck constant. In current experiments, the phonon frequency $\omega$ is tunable and can be up to about $(2\pi) 100\,{\rm kHz}$~\cite{kolkowitz2017spin}. As a result, we realize a tunable $g_k/h$ bound between $-21.6\,{\rm kHz}$ and $21.6\,{\rm kHz}$. For convenience, we fix the value of the SOC strength in real space at $g_x/h=9\,{\rm kHz}$ such that it is of the same order as $g_k/h$, which yields the atomic decay for the clock state $\kappa_e/g_x=1.1\times 10^{-8}$~\cite{kolkowitz2017spin}. We further fix the values of the decay rate and the dephasing factor at $\kappa_d/g_x=0.005$ and $\gamma_d/g_x=0.0005$, respectively. Now, the free parameters of the system reduce to phonon frequency $\omega$, Rabi frequency $\Omega$, laser detuning $\delta$, and SOC strength $g_k$. In below, we study the statistic properties of the phonon state by varying these parameters.

\begin{figure}[ptb]
	\includegraphics[width=1.0\columnwidth]{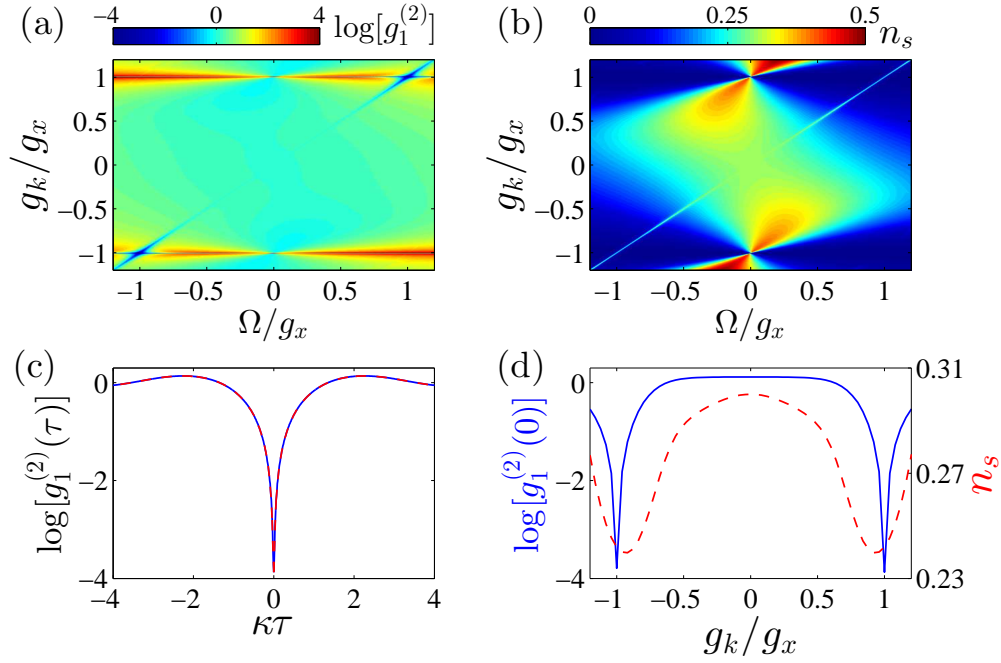}
	\caption{Distributions of $g_1^{(2)}(0)$ (a) and $n_s$ (b) on the $\Omega g_k$ parameter plane. (c) Time interval $\tau$ dependence of $g_1^{(2)}(\tau)$ for $(\Omega,g_k)=(1,1)g_x$. (d) Distributions of $g_1^{(2)}(0)$ (solid line) and $n_s$ (dashed line) along the line $g_k=\Omega$ on the $\Omega g_k$ parameter plane. In (a)-(d), the phonon frequency and the clock detuning are fixed at $\omega/g_x=1$ and $\delta/g_x=0.005$, respectively.}
	\label{PB}
\end{figure}

\section{Motional $n$-phonon bundle states} 
\begin{figure}[ptb]
	\includegraphics[width=1.0\columnwidth]{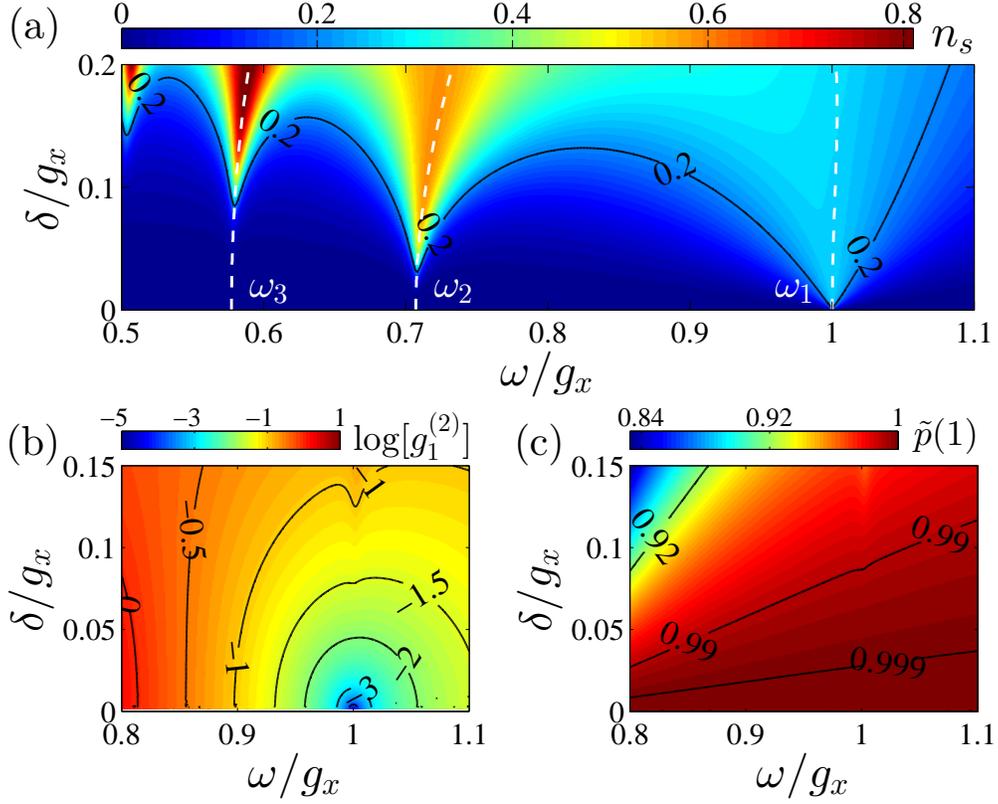}
	\caption{(a) Distribution of $n_s$ on the $\omega\delta$ parameter plane. The solid lines are the contour lines with $n_s=0.2$ and the dashed lines show the $\delta$ dependence of $\omega_n$. (b) and (c) are, respectively, the distributions of $g_1^{(2)}(0)$ and $\tilde p(1)$ on the $\omega\delta$ parameter plane around $\omega=\omega_1$. All solid lines in (b) and (c) are contour lines.}
	\label{clockshift}
\end{figure}

Before presenting our results on phonon statistics, it is instructive to explore the energy spectrum of the system. In Fig.~\ref{scheme}(b), we demonstrate the familiar level structure of a JCM ($g_k=g_x$) with $\Omega=\omega$ and $\delta=0$. The eigenenergies of the $n$th pair of dressed states $|n,\pm\rangle$ are $E_{n,\pm} = \hbar (n\omega \pm g_x\sqrt{n})$, where $n=1,2,\ldots$ and `$+$' (`$-$') denotes the upper (lower) branch. Particularly, $n$-phonon resonance occurs when the lower dressed state $|n,-\rangle$ is tuned on resonance with the vacuum state of the system, $\omega=\omega_n$, where $\omega_n = g_x/\sqrt{n}$ is the $n$-phonon resonance frequency. The energy spectrum for more general case is presented in Appendic B, where $\omega_n$ should also depend on other parameters [see, e.g. Fig.~\ref{clockshift}(a)].

Let us first consider the single phonon states by fixing the phonon frequency at $\omega=g_x$. Figure~\ref{PB}(a) and (b) display, respectively, the equal time second-order correlation function $g_1^{(2)}(0)$ and steady-state phonon number ${n}_s=\mathrm{tr}(\hat a^\dagger\hat a\rho)$ in the $\Omega g_k$ parameter plane with $\delta/g_x=0.005$. As can be seen, when $g_k$ and $\Omega$ are changed, the values of $g_1^{(2)}(0)$ and ${n}_s $ vary over a wide range. Particularly, around $(\Omega,g_k)=(\pm1,\pm1)g_x$, the system reaches the strong sub-Poissonian statistics region with $g_1^{(2)}(0)\sim 10^{-4}$. In these regions, a considerably large number of phonons (${n}_s>0.2$) are observed as well. In  Fig.~\ref{PB}(c), we further plot the interval dependence of the second-order correlation function $g_1^{(2)}(\tau)$ at $(\Omega,g_k)=(\pm1,\pm1)g_x$, which shows phonon antibunching since $g_1^{(2)}(0)<g_1^{(2)}(\tau)$. These evidences clearly show that strong phonon blockade (SPB) is achieved in these regions.

The SPB around $(\Omega,g_k)=(1,1)g_x$ can be understood by noting that the energy spectrum of the resulting JCM at $g_k=g_x$ is highly anharmonic in the strong coupling regime $g_x/\kappa\gg 1$ [Fig.~\ref{scheme}(b)]. Therefore, the condition for one phonon excitation at single-phonon resonance ($\Omega=g_x$) will block the excitation of a second phonon. At first sight, the SPB around $(\Omega,g_k)=(-1,-1)g_x$ may seem strange, as the anti-JCM realized at $g_k=-g_x$ breaks the conservation of the number of the total excitations due to the counter-rotating terms. To explain this, we note that an anti-JCM is equivalent to a JCM under the unitary transformation $e^{-i\pi\sigma_x/2}$, i.e., $\sigma_\pm\rightarrow\sigma_\mp$ and $\sigma_z\rightarrow-\sigma_z$. Consequently, a SPB should occur at $-\Omega=g_x$. We point out that, because of the small phonon decay in the clock transition, the lifetime of the phonon blockade ($\tau\sim \kappa_d^{-1}$) in our system can be very long.

To reveal more details, we also plot, in Fig.~\ref{PB}(d), the distributions of $g_1^{(2)}(0)$ and $n_s$ along the line $g_k=\Omega$ on the $\Omega g_k$ plane. As can be seen, both $g_1^{(2)}(0)$ and $n_s$ possess two local minima close to $g_k=\pm g_x$. In addition, $g_1^{(2)}(0)$ grows rapidly from these minima when $g_k$ is slightly tuned away from $\pm g_x$. And $g_1^{(2)}(0)$ approaches unity for a coherent phonon state at $g_k=0$ where Hamiltonian~(\ref{Rabi1}) reduces to a standard QRM.

We now turn to study the properties of phonon states in the atom-phonon resonance regime, i.e., $\Omega=\omega$. In addition, due to the equivalence of two SPB regions in Fig.~\ref{PB}(a), we shall focus, without loss of generality, on the SPB region with $g_k=g_x$. The atom-phonon resonance condition can be satisfied by tuning external atom trap potential and/or power of the classical laser field in the experiment. In particular, it should be noted that our results remain qualitatively unchanged even deviating from the resonance condition, corresponding the slightly shift the position of $n$-phonon resonance $\omega_n$.

Figure~\ref{clockshift}(a) shows the phonon number $n_s$ as a function of the phonon frequency $\omega$ and clock shift $\delta$. Particularly, the dashed lines plot the $\delta$ dependence of the $n$-phonon resonance frequency $\omega_n$ obtained by numerically diagonalizing ${\mathcal H}'$. An immediate observation is that, for a given $\delta$, $n_s(\omega)$ exhibits multiple peaks at exactly the phonon resonance frequencies $\omega_n$, signaling the existence of possible multiphonon states. Remarkably, because our scheme operates in the resonant regime, the average phonon number of the phonon states generated here is much larger than that of the phonon states produced in the dispersive regime. In addition, there exists a large area on the $\omega\delta$ plane in which the average phonon number is higher than $0.2$.

To explore the statistic properties of the phonon emissions, we first map out, in Fig.~\ref{clockshift}(b), $g_1^{(2)}(0)$ on the $\omega\delta$ parameter plane around $\omega=\omega_1$. As can be seen, phonon blockade is realized for the whole parameter space covered by Fig.~\ref{clockshift}(b). There even exists a large area on the $\omega\delta$ plane such that the strong phonon blockade condition, say $g_1^{(2)}(0)<10^{-2}$, is satisfied. Then combined with requirement of large phonon emission number, say $n_s>0.2$, our system can be used as a high-quality single motional phonon source operating in a parameter regime that is easily accessible to current experiments. To further quantify the quality of the single-phonon states around $\omega_1$, we introduce $\tilde p(q)\equiv qp(q)/n_s$ which measures the fraction of $q$-phonon states among the total emitted phonons. In Fig.~\ref{clockshift}(c), we plot the distribution of $\tilde p(1)$ on the $\omega\delta$ plane. As can be seen, for the parameter region of our interest, nearly 100\% of phonon emission is of the single-phonon nature. It should also be noted that $\tilde p(1)$ alone is not sufficient to judge the quality of the single-phonon source, because, by comparing Fig.~\ref{clockshift}(b) and (c), the main feature of $\tilde p(1)$ is inconsistent with that of $g_1^{(2)}(0)$ when $n_s$ is small.

\begin{figure}[ptb]
	\includegraphics[width=1.0\columnwidth]{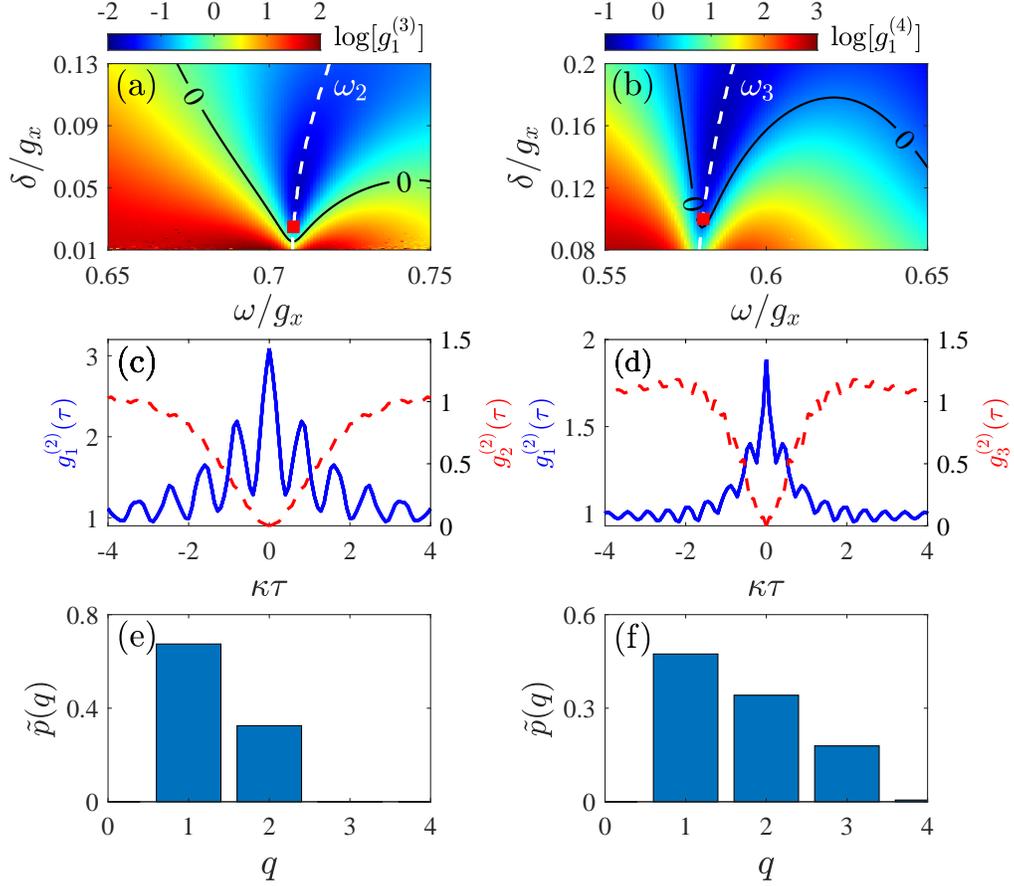}
	\caption{Statistic properties of motional two-phonon (left column) and three-phonon (right column) states. (a) and (b) show the distributions of $g_1^{(3)}(0)$ and $g_1^{(4)}(0)$, respectively, on the $\omega\delta$ parameter plane. Solid lines in (a) and (b) are contour lines marking $g_1^{(3)}(0)=1$ and $g_1^{(4)}(0)=1$, respectively. The dashed lines in (a) and (b) represent the $\delta$ dependence of $\omega_2$ and $\omega_3$, respectively. (c) plots $g_1^{(2)}(\tau)$ (solid line) and $g_2^{(2)}(\tau)$ (dashed line) for $(\delta,\omega)=(0.025,0.708)g_x$, i.e., the red square in (a). (d) shows $g_1^{(2)}(\tau)$ (solid line) and $g_3^{(2)}(\tau)$ (dashed line) for $(\delta,\omega)=(0.1,0.58)g_x$, i.e., the red square in (b). (e) and (f) plot the steady-state phonon-number distribution $\tilde p(q)$ corresponding to parameters marked by red squares in (a) and (b), respectively.}
	\label{g3_g4}
\end{figure}

Finally, we study the properties of the emitted phonon in the multiphonon resonance regime. As shown in Fig.~\ref{clockshift}(a), only for sufficiently large pumping, i.e., $\delta/g\geq0.015$ and $0.094$ for two- and three-phonon resonances respectively, the phonon number $n_s$ around $\omega_2$ and $\omega_3$ become significant ($>0.2$). We shall then explore the statistic properties of the multiphonon resonances in these regions. Figure~\ref{g3_g4} summarizes the main properties of the motional multiphonon states. Due to the similarity between the two- and three-phonon states, we shall only discuss the two-phonon emissions. In Fig.~\ref{g3_g4}(a), we map out the third-order correlation function in the two-phonon resonance regime of the $\omega\delta$ plane. As can be seen, there exist a large parameter regime with $g_1^{(3)}(0)<1$ in which the $(n+1)$-phonon emission is blockades. Meanwhile, the minimum value of $g_1^{(3)}(0)$ is achieved at $\omega=\omega_2$, where the highest phonon numbers for the multiphonon resonance is reached. This indicates that at the strongest three-phonon blockade, we have the highest phonon number.

To further confirm the bundle-emission nature of the phonon states, we plot, in Fig.~\ref{g3_g4}(c), the typical interval dependence of the correlation functions $g_1^{(2)}(\tau)$ and $g_{2}^{(2)}(\tau)$. As can be seen, the criterions for motional two-phonon bundle states, $g_1^{(2)}(0)>g_1^{(2)}(\tau)$ and $g_2^{(2)}(0)<g_2^{(2)}(\tau)$, are indeed satisfied. Another observation from Fig.~\ref{g3_g4}(c) is that the decay times for both $g_1^{(2)}(\tau)$ and $g_2^{(2)}(\tau)$ are proportional to $\kappa_d^{-1}$, which indicates that the decay of the bunching for single phonon and the decay of the antibunching for the separated bundles of phonons are of the same timescale for the two-phonon bundle states. The two-phonon nature of the emission is also demonstrated by the phonon-number distribution $\tilde p(q)$ shown in Fig.~\ref{g3_g4}(e). Indeed, for the two-phonon emission case, $\tilde p(q)$ becomes negligibly small for $q>2$. We point out that, as shown in the right panels of Fig.~\ref{g3_g4}, the three-phonon emission possesses similar statistic properties as those of two-phonon emission. The observed phonon probability for steady-state $n$-phonon bundles states exhibits a monotonical decreasing distribution with increasing $q$ for $1\leq  q \leq n$, arising in a dynamical processes of bundle emissions~\cite{munoz2014emitters}. In general, the $n$-phonon bundle states contains the various dynamical processes of emissions in Fock state $|q\rangle$ with a distinguishable short temporal window $\sim 1/(q\kappa_d)$. Thus the multiphonon state ($|q\rangle$) can be directly extracted with a high phonon probability by choosing a very short temporal window~\cite{munoz2014emitters,PhysRevLett.124.053601}. Moreover, we should note that the generated motional $n$-phonon bundle states is essential different from the experimental observed $n$-quanta blockade, i.e., two-photon blockade~\cite{PhysRevLett.118.133604}, where the quantum statistic for the latter only characterizes the single photon but not for separated bundles of photons with satisfying $n$-photon bunching $g_1^{(n)}(0)>1$ and  $(n+1)$-photon antibunching $g_1^{(n+1)}(0)<1$ as well. 

We remark that the underlying reason that $n$-phonon bundle states are well-resolved in the parameters space is due to the strong anharmonicity of the energy spectrum such that the condition $|\omega_n-\omega_m|\gg \kappa_d$ is satisfied for any $n\neq m$. This is in striking contrast to the proposed $n$-photon bundle emission utilizing Mollow physics~\cite{mollow1969power} in which the $(n+1)$th-order process of quantum states is used~\cite{munoz2014emitters,munoz2018filtering}. In addition, the whole process must be operated in the far dispersive regime and under a strong pump field, which, in terms of our model, requires $|\omega-\Omega|\gg g_x\sqrt{n+1}$ and $|\delta|\gg g_x\sqrt{n+1}$, simultaneously. Finally, we
check that our numerical results do not affect when includes the weak atomic decay of clock state ($\kappa_e/\kappa_d\ll 1$) and phonon dephasing ($\gamma_d/\kappa_d\ll 1$), albeit the strong damping of $\gamma_d$ and $\gamma_e$ could induce a significant decoherence for realization of nonclassical quantum states~\cite{PhysRevB.81.245419,PhysRevLett.104.073904,PhysRevLett.108.093604}.

\section{Conclusions}
\noindent
Based upon the currently availably techniques in experiments, we have proposed to generate motional $n$-phonon bundle states using a trapped alkaline-earth atom driven by a clock laser. Since our system works in the resonant regime, the steady-state phonon number containing in the bundle states is three orders of magnitude larger than those obtained in the earlier theoretical schemes operating in the far dispersive regime. Moreover, the quality of the $n$-phonon bundle states is also demonstrated by the strong antibunching for the separated bundles of phonons and bunching for single phonon. Finally, we emphasize that the nonclassical nature of the long-lived motional $n$-phonon bundle states can be quantum-state transferred to photonic mode by applying a readout cavity field with phonon-photon beam-splitter interaction~\cite{PhysRevLett.107.133601,PhysRevLett.112.143601}. Moreover, as the scheme mitigates the laser induced heating, it could be inspired an interesting opportunity of exploration Dicke phase transition for the external motional modes~\cite{hamner2014dicke}, superradiances from the clock transition~\cite{norcia2018frequency} and novel quantum states of matters hindered by heating~\cite{PhysRevX.6.031022,zhou2017symmetry,iemini2017majorana}. Our proposal for trapped single atom could be equivalently applied to the hybrid spin-mechanical systems~\cite{PhysRevLett.125.153602,PhysRevA.98.052346}. In particular, the parameter in our model is outside the Lamb-Dicke regime with $|\kappa x_0|\sim 0.2$, which could provide a versatile platform for exploring long-lived mesoscopic entanglement for trapped atom~\cite{PhysRevLett.98.063603}. Furthermore, we could expect that the proposed system provides versatile applications in quantum metrology limited by decoherence~\cite{PhysRevLett.76.1796,wolf2019motional} and in fundamental tests of quantum physics~\cite{kozlov2018highly}.

\setcounter{equation}{0}
\setcounter{figure}{0}
\renewcommand{\theequation}{S\arabic{equation}}
\renewcommand{\thefigure}{S\arabic{figure}}
\section*{APPENDIX A:  MODEL HAMILTONIAN}\label{sec:Ham}
\noindent
We present the derivation of the generalized quantum Rabi model (QRM) of phonon by utilizing optical clock transition in in ultracold single atom. For specificity, we consider an optical clock transition frequency of $^1 S_0- ^3P_0$ is $\omega_a$ in a single ultracold $^{87}$Sr atom, which including a ground ground-state $|g\rangle$ and an excited-state $|e\rangle$. Here the $|g\rangle = ^{1}$$S_0$ is the ground state and $|e\rangle = ^3$$P_0$ is an exceptionally long-lived electronic state (160 seconds). The single atom is resonant coupled by a linearly $\pi$-polarized classical plane wave laser with the frequency $\omega_c$ and wavelength $\lambda_C=698$ nm, which is propagating in the $xy$ plane making an angle $\phi$ to the $x$ axis. As a result, the spontaneous emission and decoherence of excited state can be safely ignored, which is of paramount importance to realizations of motional $n$-phonon states in our model.

In addition, the single two-level atom is confined in a spin dependent one-dimensiona harmonic trap $V(x)= \frac{1}{2}M \omega^2 x^2$ generated by a $\pi$-polarized laser at the `magic' wavelength $\lambda_L=813$ nm, where $M$ is the mass of atom and $\omega$ is the trap frequency. Now, it can be read out that the Hamiltonian for the internal states of an atom under the rotating-wave approximation reads
\begin{align}
	{\cal H}/\hbar  &= \frac{{\mathbf p}^{2}}{2M}  + \Omega_l(x) \sigma_{-} + \Omega^*_l(x) \sigma_{+} + {\delta}\sigma_z +V(x)  \label{SM-single}
\end{align}
where $\Omega_l(x) = (\Omega_0x-i{\Omega}/{2}) e^{-i\kappa x}$ is the spatial dependent Rabi frequency with $\Omega_0$ and $\Omega$ being the coupling strengths, $\delta=\omega_a-\omega_c$ is the highly tunable single-photon detuning of clock transition, and $\kappa=k_{C}\cos\phi$ is the effective tunable laser wave vector corresponding a tilted angle $\phi$ respect to the $x$ axis. Here, $\sigma_{x,y,z}$ are Pauli matrices for spin-$1/2$ system with $\sigma_\pm=(\sigma_x\pm i\sigma_y)/2$. It is easy to check that these operators satisfy the commutation relations:

After performing the gauge transformation $\left|g\right\rangle \rightarrow e^{i\kappa x/2}\left|g\right\rangle$ and $\left|e\right\rangle \rightarrow e^{-i\kappa x/2}\left|e\right\rangle$~\cite{PhysRevLett.108.125301} for eliminating the prefactor $ e^{\pm i\kappa x}$ in the spin-flip Raman term, the single-particle Hamiltonian of Eq.~(\ref{SM-single}) is given by (with ignoring the constant term)
\begin{eqnarray}
	{\cal{H}}/\hbar &= &\frac{({\mathbf{p}}-{\mathbf{A}})^{2}}{2M}+   (\Omega_0x-i{\Omega}/{2})  \sigma_{-} +  (\Omega_0x+i{\Omega}/{2}) \sigma_{+}  + V(x), \nonumber \\ 
	&= &\frac{({\mathbf{p}}-{\mathbf{A}})^{2}}{2M}+ \Omega_0x \sigma_x + \frac{\Omega}{2}\sigma_y + {\delta}\sigma_z + V(x), \nonumber \\
	&\equiv & \frac{{\mathbf{p}}^{2}}{2M} + \kappa_{\rm{so}} p_x \sigma_z + \Omega_0x \sigma_x + \frac{\Omega}{2}\sigma_y + {\delta}\sigma_z+ V(x)
	\label{SM-single-SO}
\end{eqnarray}
where ${\mathbf{A}}=-\hbar\kappa\sigma_z/2$ is the vector potential through Raman processe~\cite{NATYJ2011SOCAT}, $\kappa_{\rm{so}}=\hbar \kappa/2M$ and $\Omega_0$ characterizes the strength of the 1D spin-orbit (SO) coupling in real and momentum space, respectively. In particular, the position-dependent Raman term of $\Omega_0x \sigma_x$ is equivalently to applying a  spatially dependent gradient magnetic field along $x$-axis, which can be experimentally generated by tailoring the clock laser by using a spatial light modulator~\cite{Beeler,Palima:07,Pasienski:08,PMID:23301155,Gaunt12}. We should note that this linear spatial dependence Rabi frequency has been experimentally realized for studying the spin Hall effect of ultracold quantum gases in Ref.~\cite{Beeler}. As can be seen, the spatial dependent Rabi coupling $\Omega_l(x)$ plays an essential role in generation SO couplings in both real and momentum spaces, simultaneously.

To gain more insight, we introduce the position-momentum representation, $x=\frac{1}{\sqrt{2}}\sqrt{\frac{\hbar}{M\omega}}(\hat a^\dag + \hat
a)$, $p_x=\frac{i}{\sqrt{2}}\sqrt{M\hbar\omega}(\hat a^\dag - \hat
a)$ with $\hat{a}$ denoting the annilation operator of bosonic phonon mode with harmonic oscillator frequency $\omega$. Then the Hamiltonian of equation (\ref{SM-single-SO}) can be rewritten as
\begin{align}
	{\cal {H}}/\hbar  \!=\! \omega\hat{a}^\dag\hat{a} \!+\! \frac{\Omega}{2}\sigma_y \!+\! {\delta}\sigma_z\!+\! \frac{g_x}{2} (\hat{a}^\dag \!+\! \hat{a})\sigma_x \!+\! \frac{ig_{k}}{2}(\hat{a}^\dag \!-\!\hat{a})\sigma_z,
	\label{SM-Rabi}
\end{align}
where the single-phonon atom coupling strengths $g_x=2\Omega_0 x_0$ and $g_k=\kappa\omega x_0$ are emerged by the two-type SO coupling in real and momentum spaces, corresponding $x_0 = \sqrt{{\hbar}/({2M\omega})}$ being the  zero-point fluctuation amplitude of trapped atom oscillator. Finally, we obtain the tunable generalized QRM Hamiltonian (1) in the main text.

\section*{APPENDIX B: ENERGY SPECTRA for GENERALIZED QRM}\label{sec:spe}
\noindent
We present the details on the derivation of the energy spectrum for the generalized QRM. By introduce a gauge transformation of the spin rotation $R_x= e^{{i\pi}\sigma_x/4}$, the Hamiltonian (\ref{SM-Rabi}) reduces to
\begin{align}
	{\cal {H}}/\hbar &= \omega\hat{a}^\dag\hat{a} \!+\! \frac{\Omega}{2}\sigma_z \!-\! {\delta}\sigma_y\!+\! \frac{g_x}{2} (\hat{a}^\dag \!+\! \hat{a})\sigma_x \!-\! \frac{ig_{k}}{2}(\hat{a}^\dag \!\!-\!\hat{a})\sigma_y \nonumber \\
	&= \omega\hat{a}^\dag\hat{a} \!+\! \frac{\Omega}{2}\sigma_z \!-\! {\delta}\sigma_y\!+\! \lambda_+(\hat{a}^\dag \sigma_- + \hat{a}\sigma_+) + \lambda_-(\hat{a}^\dag \sigma_+ + \hat{a}\sigma_-)
	\label{SM-Rabi1}
\end{align}
corresponding the unitary transformation $R_x^\dag \sigma_z R_x = - \sigma_y$ and $R_x^\dag \sigma_y R_x=  \sigma_z$. Here  $\lambda_\pm=(g_x\pm g_{k})/2$ are introduced for shorthand notation. For fixing $\delta=0$, the Eq.~(\ref{SM-Rabi1}) satisfies the parity with $[{\cal H},{\cal P}]=0$, where the parity operator ${\cal P} = \exp[i\pi(\hat{a}^\dag\hat{a}+(1+\sigma_z)/2)]$ measures an even-odd parity of total excitation number. Note that the Hamiltonian terms proportional to $\hat{a}\sigma_{-}$ (counter-rotating wave coupling terms) does not conserve the number of bare excitations. However since describe the simultaneous creation or destruction of two excitations, the parity of the excitation number is conserved.

\begin{figure}[ptb]
	\includegraphics[width=0.9\columnwidth]{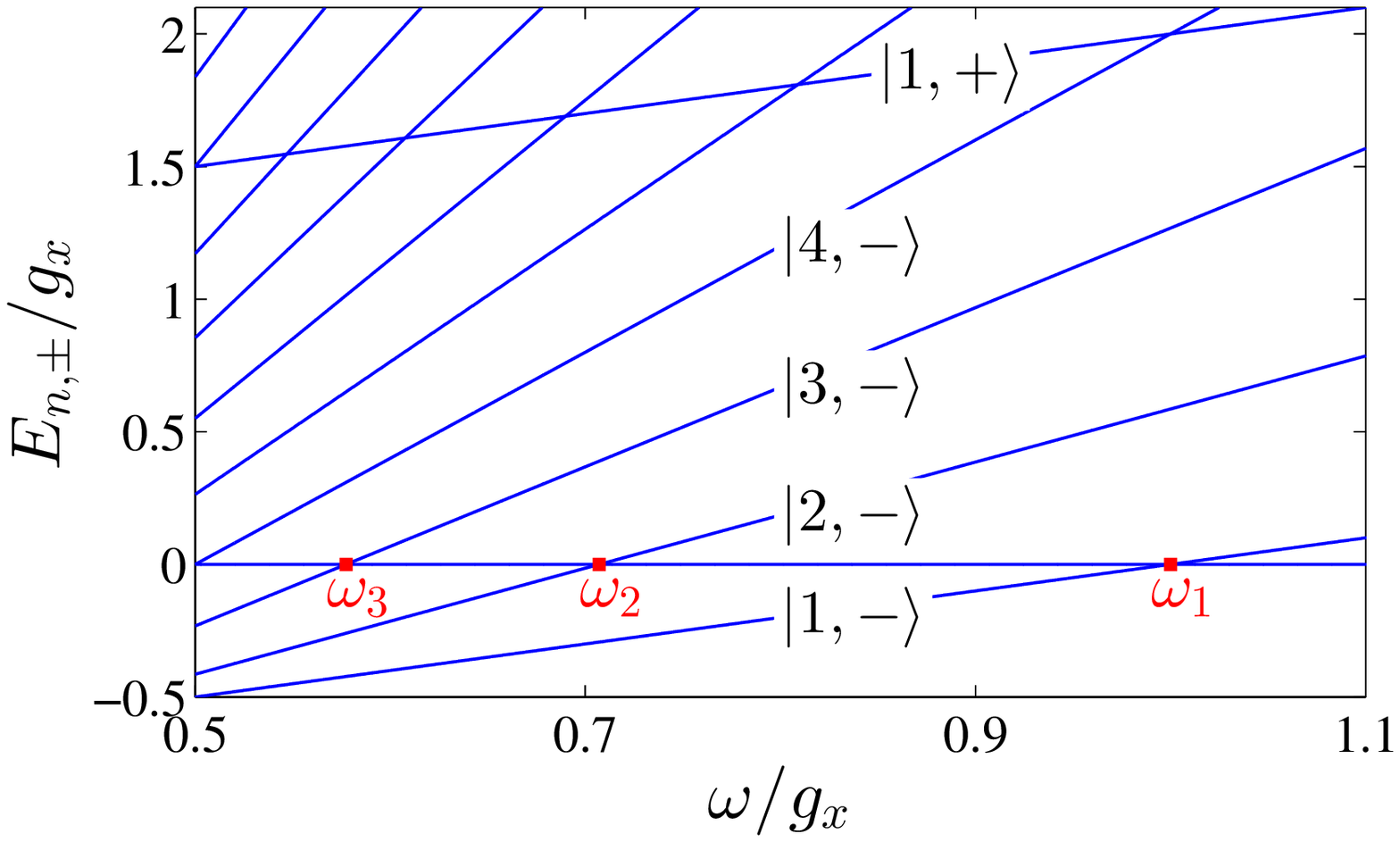}
	\caption{The typical energy spectrum of Hamiltonian  (\ref{SM-Rabi1}) with the zero clock shift $\delta=0$. The red square shows the position of $n$-phonon resonance $\omega_n$. And the $|n,\pm\rangle =[|n,g\rangle \mp |n-1,e\rangle ]\sqrt{2}$ denotes the $n$th pair of dressed states of the system. The other parameters are $g_k/g_x=1$ and $\Omega/\omega=1$.}
	\label{SM_spe}
\end{figure}

\begin{figure}[ptb]
	\includegraphics[width=0.9\columnwidth]{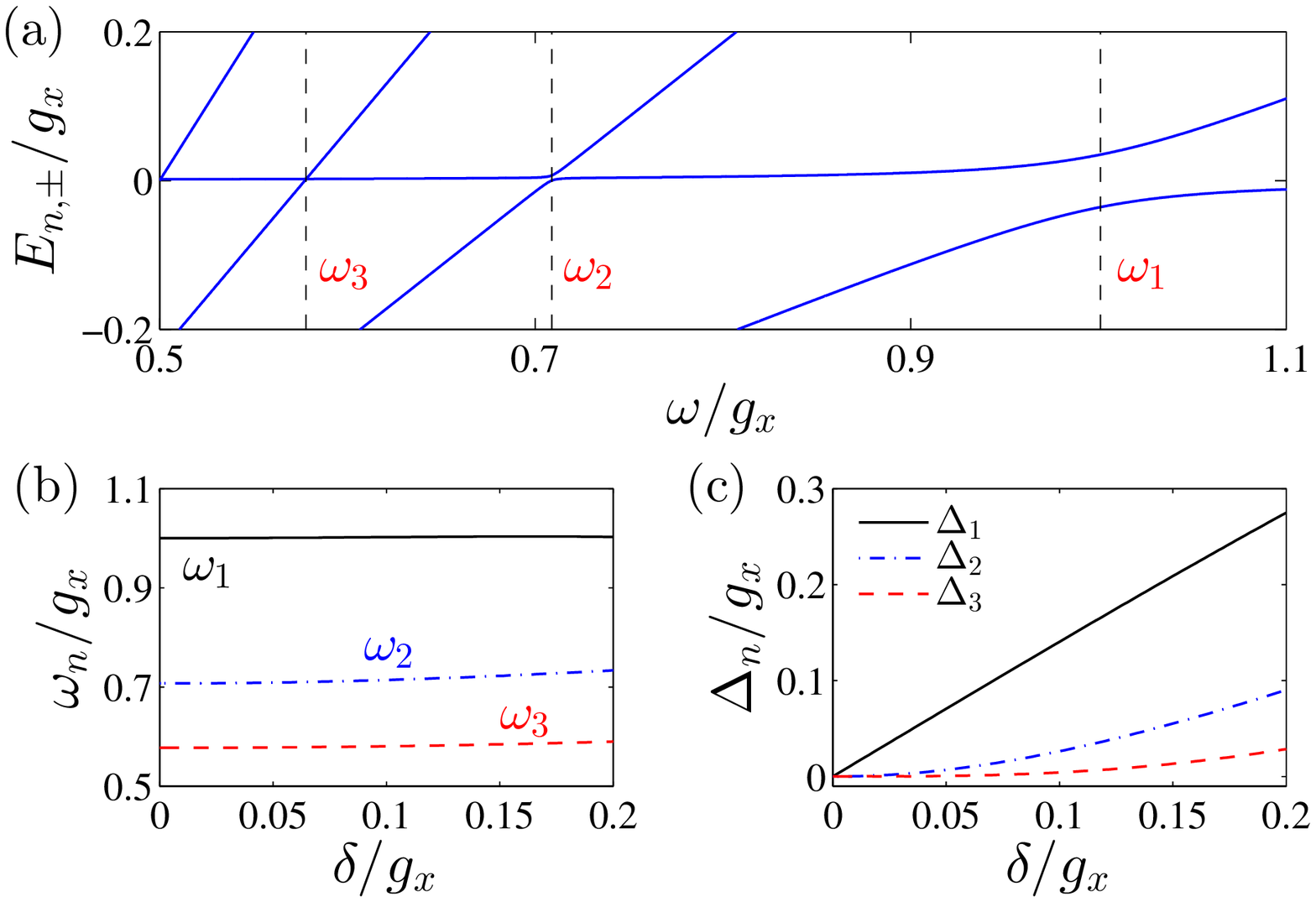}
	\caption{(a) The energy spectrum of Hamiltonian  (\ref{SM-Rabi1}) with $\delta/g_x=0.05$. The vertical dashed lines denoting the positions of $\omega_{n}$ are guides for the eyes. The $\delta$ dependence of $\omega_n$ (b) and $\Delta_n$ (c) for the $n$-phonon states, respectively. The other parameters are $g_k/g_x=1$ and $\Omega/\omega=1$.}
	\label{SM_gap}
\end{figure}

For fixing $g_k= g_x$ and $\delta=0$, the Hamiltonian~(\ref{SM-Rabi1}) can be reduced to a seminal Jaynes-Cummings model whose Hamiltonian only possesses the rotating terms with hosting the number of total excitation. As a result, the energy eigenvalues for the $n$th pair of dressed states satisfy $|n,\pm\rangle =[|n,g\rangle \mp |n-1,e\rangle ]\sqrt{2}$ with $\omega/\Omega=1$, corresponding the energy eigenvalues $E_{n,\pm} = \hbar (n\omega \pm g_x\sqrt{n})$. Here $n$ is the phonon emissions number and $+$ ($-$) denotes the higher (lower) branch. We find that the dressed states splittings of $|n,\pm\rangle$ are $\pm g_x\sqrt{n}$, corresponding the typical anharmonic energy spectrum, as shown in Fig.~\ref{SM_spe}. In addition, the $n$-phonon resonance occurs at $\omega_n=g_x/\sqrt{n}$ ($\omega_n>0$) when the lower dressed state $|n,-\rangle$ is tuned resonant with the vacuum state of the system.

In the presence of $\delta$, the Hamiltonian of Eq.~(\ref{SM-Rabi1}) breaks ${\cal P}$-symmetry and lacks of the analytical solution. Without loss of generality, we may expand the wavefunction of the system that is formally given as
\begin{align}
	|\psi\rangle &= \sum_{n=0}^{\infty} C_{n,g} |n, g\rangle
	+\sum_{n=0}^{\infty} C_{n,e}|n,e\rangle,
\end{align}
where a set of eigenstates $|n, g\rangle=|n\rangle \otimes |g\rangle$ and $|n, e\rangle=|n\rangle \otimes |e\rangle$ are the product states of the two-level atomic states and Fock states of phonon, and $|C_{n,g}|^2$ and $|C_{n,e}|^2$ denote the atomic occupation probability for eigenstates $|n, g\rangle$ and $|n, e\rangle$, respectively. Thus the Schr\"{o}dinger equation of the QRM Hamiltonian (\ref{SM-Rabi1}) reads $i \frac{d |\psi\rangle }{dt} = {\cal H} |\psi\rangle$,
which yields
\begin{align}
	i \frac{d |\psi\rangle }{dt} &= \sum_{n=0}^{\infty} \left(\dot{C}_{n,g} |n, g\rangle
	+\dot{C}_{n, e}|n,e\rangle\right), \\
	{\cal H}|{\psi}\rangle &= \sum_{n=0}^{\infty} \left( n\omega {C}_{n,g} |n, g\rangle
	+(n\omega +\Omega){C}_{n, e}|n,e\rangle \right), \nonumber \\
	&= \sum_{n=0}^{\infty} \left( i\delta{C}_{n,e} |n, g\rangle -i\delta{C}_{n, g}|n,e\rangle \right), \nonumber \\
	& + \sum_{n=0}^{\infty} \left( \sqrt{n}\lambda_+{C}_{n,g} |n-1, e\rangle
	+\sqrt{n+1}\lambda_+{C}_{n,e} |n+1, g\rangle \right), \nonumber \\
	& + \sum_{n=0}^{\infty} \left( \sqrt{n+1}\lambda_-{C}_{n,g} |n+1, e\rangle
	+\sqrt{n}\lambda_-{C}_{n,e} |n-1, g\rangle \right),
\end{align}
leading to the evolution equations are given by
\begin{align}
	\dot{C}_{n,g} &=  n\omega {C}_{n,g}
	+ i\delta{C}_{n,e}
	+\sqrt{n}\lambda_+{C}_{n-1,e}
	+\sqrt{n+1}\lambda_-{C}_{n+1,e}, \\
	\dot{C}_{n, e} &=(n\omega +\Omega){C}_{n,e}-i\delta{C}_{n, g} + \sqrt{n+1}\lambda_+{C}_{n+1,g}
	+ \sqrt{n}\lambda_-{C}_{n-1,g}
	\label{SM-evo}
\end{align}
As a result, the QRM Hamiltonian decouples into an infinite direct product of $2\times n$-matrix Hamiltonian
\begin{align}
	\left(\begin{array}{cccccccc}
		\ddots & \vdots & \vdots & \vdots & \vdots & \vdots & \vdots & \rotatebox[]{90}{$\ddots$}\\
		\cdots &  n\omega & i\delta & 0 & \sqrt{n+1}\lambda_- & 0 & 0  & \cdots\\
		\cdots & -i\delta & n\omega +\Omega & \sqrt{n+1}\lambda_+ & 0 & 0 & 0 & \cdots \\
		\cdots & 0 & \sqrt{n+1}\lambda_+ &  (n+1)\omega & i\delta & 0 & \sqrt{n+2}\lambda_-  & \cdots\\
		\cdots & \sqrt{n+1}\lambda_-  &  0 & -i\delta & (n+1)\omega+\Omega & \sqrt{n+2}\lambda_+ & 0 & \cdots \\
		\cdots & 0 & 0 & 0 & \sqrt{n+2}\lambda_+ & (n+2)\omega & i\delta & \cdots\\
		\cdots & 0 & 0 & \sqrt{n+2}\lambda_- & 0 & -i\delta & (n+2)\omega+\Omega & \cdots \\
		\rotatebox[]{90}{$\ddots$} & \vdots & \vdots & \vdots & \vdots & \vdots & \vdots & \ddots
	\end{array}\right)
	\label{SM-spec}
\end{align}
In general, the matrix equation (\ref{SM-spec}) can be solved numerically to yield the energy spectrum of the system.

Figure~\ref{SM_gap} shows the typical energy spectrum in the presence of the clock shift $\delta/g_x=0.05$. As can be seen, the level structure is slightly distorted for weak enough clock detuning. In particular, the energy spectrum at the $n$-phonon transition ($\omega=\omega_n$) becomes anticrossing with opening the gap $\Delta_n$ when $\delta\neq 0$. To further characterize the energy spectrum, we plot the $n$-phonon resonance $\omega_n$ and corresponding the energy gap $\Delta_n$ as a function of $\delta$ as shown in Figs.~\ref{SM_gap}(b) and \ref{SM_gap}(c), respectively.  Although the $\omega_n$ slightly respects to $\delta$, the $\Delta_n$ is increasing with increasing $\delta$. Therefore, the clock shift $\delta$ can be used as a control knob for realizing motional $n$-phonon bundle states, such that it plays a significant role for the phonon number emissions and quantum statistics, as shown in the main text.

\begin{figure}[ptb]
	\includegraphics[width=0.9\columnwidth]{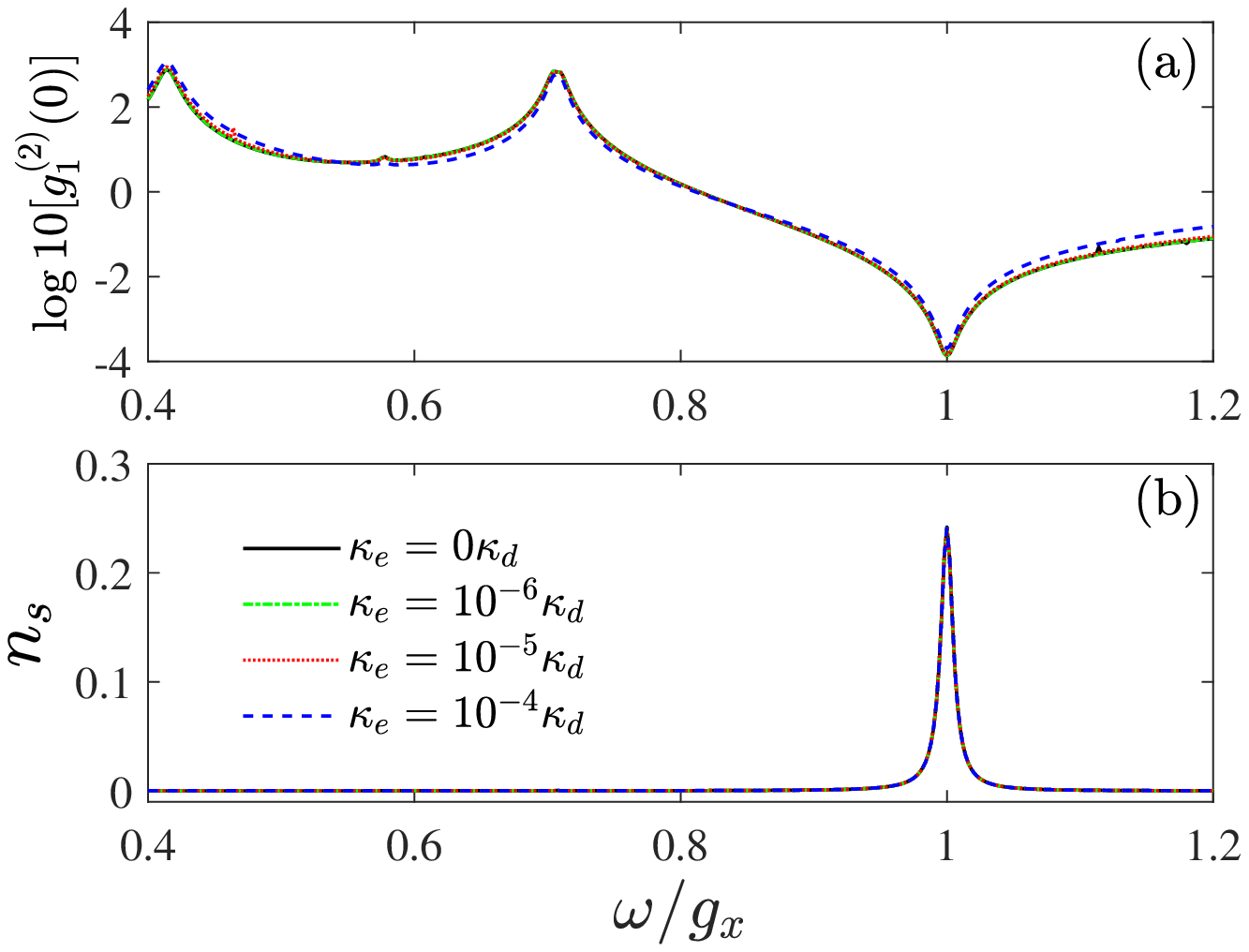}
	\caption{ The $\omega$ dependence of $g_1^{(2)}(0)$ in (a) and $n_s$ in (b) for different value of $\kappa_e$ with $\delta/g_x=0.005$ and $\gamma_d/\kappa_d=0.1$. }
	\label{gamma}
\end{figure}

\begin{figure}[ptb]
	\includegraphics[width=0.9\columnwidth]{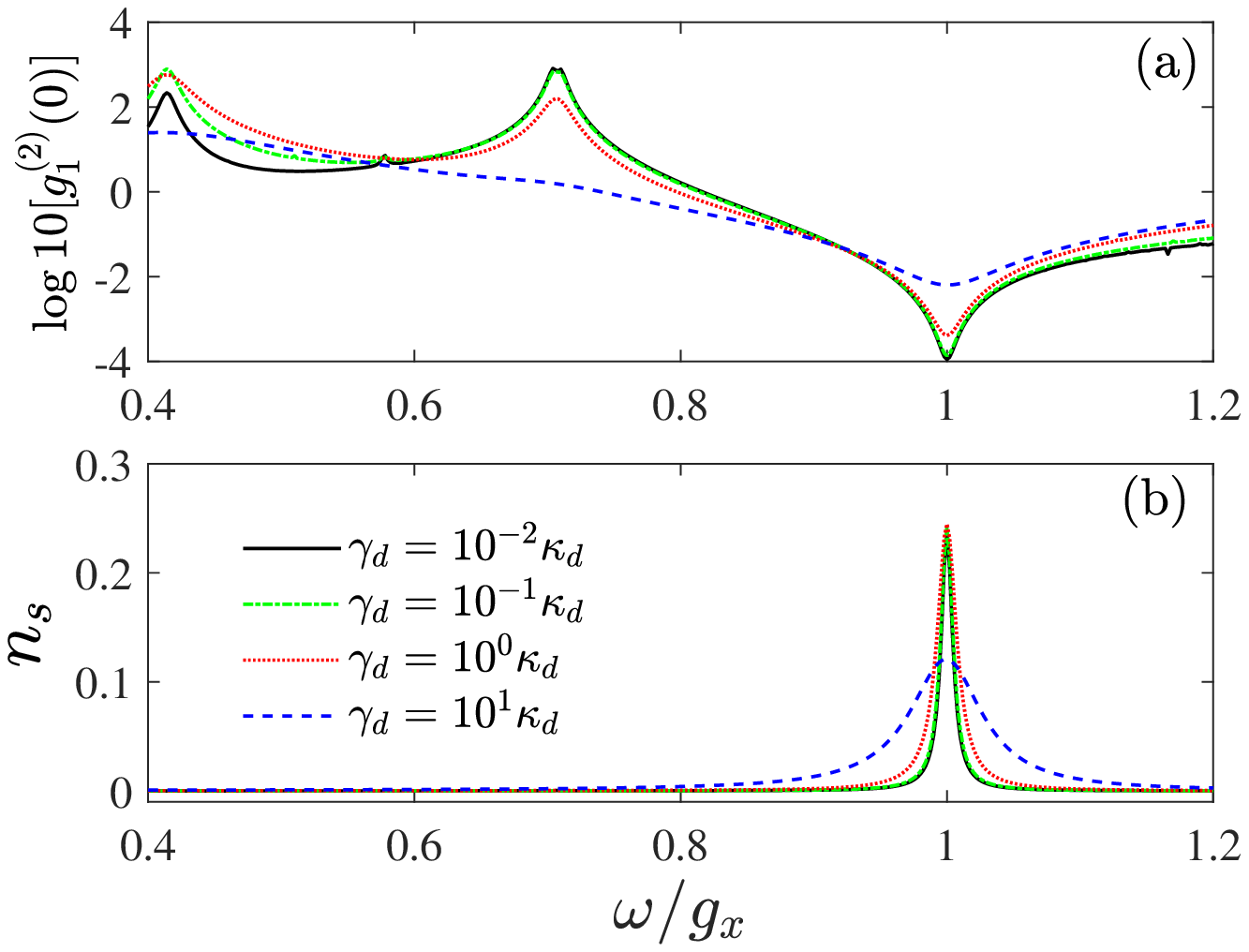}
	\caption{ The $\omega$ dependence of $g_1^{(2)}(0)$ in (a) and $n_s$ in (b) for different value of $\gamma_d$ with $\delta/g_x=0.005$ and $\kappa_e/\kappa_d=2.2\times 10^{-6}$. }
	\label{dephasing}
\end{figure}

\section*{APPENDIX C: THE EFFECT of DECAY and DEPHASING}\label{sec:damping}
\noindent
We present the numerical results of the antibunching of phonon with different atomic decay and dephasing. Figure~\ref{gamma}(a) and \ref{gamma}(b) shows the second-order correlation function $g_1^{(2)}(0)$ and corresponding phonon number $n_s$ as a function of the phonon frequency $\omega$ for different values of atomic decay for clock state $\gamma_e$, respectively. As can be seen, both $g_1^{(2)}(0)$ and $n_s$ are immune to the weak atomic decay when $\kappa_e/\kappa_d\ll 1$. We should note that the atomic decay is very small with $\gamma_e/\kappa=2.2\times 10^{-6}$ by employed the advantages of energy-level structures in alkaline-earth-metal $^{87}$Sr atom, which can be safely ignored in our numerical simulation.

To proceed further, we investigate the effect of dephasing $\gamma_d$ on generation of phonon blockade, as shown in Fig.~\ref{dephasing}. For large $\gamma_d$, it's clear that dephasing will play an important role in quantum system including the phonon excitation $n_s$ and quantum statistic $g_1^{(2)}(0)$. With increasing $\gamma_d$, the antibunching $g_1^{(2)}(0)$ is rapidly growing, albeit $n_s$ is not insensitive to the moderate dephasing with $\gamma_d/\kappa_d <1$. In particular, the strong phonon blockade will be broken with the antibunching  $g_1^{(2)}(0)>0.01$ in the present of large dephasing rate $\gamma_d/\kappa_d\gg1$. We also checked that the large dephasing $\gamma_d$ could induce a significant decoherence for realization high-quality of motional $n$-phonon bundle states except for single phonon blockade.

\section*{Funding}
\noindent
We are grateful to Yue Chang for insightful discussions. This work was supported by the National Key R$\&$D Program of China (Grant No. 2018YFA0307500 and  No. 2017YFA0304501), NSFC (Grant No. 11874433, No. 11674334, No. 11974363, and No. 11947302), and the Key-Area Research and Development Program of GuangDong Province under Grants No. 2019B030330001.



\section*{Disclosures}
\noindent
The authors declare no conflicts of interest.

\end{document}